\documentclass[reprint, floatfix]{revtex4-1}
\usepackage[utf8]{inputenc}
\usepackage{graphicx,amssymb} %
\usepackage{romannum}
\usepackage{xspace}
\usepackage{mathtools}
\usepackage{tikz}
\usetikzlibrary{calc,arrows}
\usetikzlibrary{trees}
\usetikzlibrary{shapes}
\usetikzlibrary{positioning,shapes,shadows,arrows}
\usetikzlibrary{decorations.pathmorphing}
\usetikzlibrary{decorations.markings}
\usetikzlibrary{decorations.pathreplacing}
\usetikzlibrary{automata,positioning}
\tikzset{
    photon/.style={decorate, decoration={snake,pre length=2pt,post length=1pt, amplitude=1.5pt, segment length=4pt}, draw=red},
    photonarr/.style={draw=red, decoration={snake}, decorate, postaction={decoration={markings, mark=at position 1 with {\arrow[draw=red]{>}}}, decorate}},
    flash/.style={dashed, draw=red, postaction={decorate},
        decoration={markings,mark=at position 1 with {\arrow[draw=red]{>}}}},
	lepton/.style={draw=blue, postaction={decorate},
        decoration={markings,mark=at position .55 with {\arrow[draw=blue]{>}}}},
    nucleon/.style={draw=black, postaction={decorate},
        decoration={markings,mark=at position .55 with {\arrow[draw=black]{>}}}},
	nucleonsimple/.style={draw=black, double distance=4pt, postaction={decorate},
		decoration={markings,mark=at position .55 with {\arrow[draw=black, line width=0.3, scale=5]{>}}}},
    gluon/.style={decorate, draw=black,
        decoration={coil,amplitude=1.5pt, segment length=2pt}},
}

\newcommand{\RM}[1]{\MakeUppercase{\romannumeral #1}}
\newcommand{\COMPASS}{\textsc{Compass}\xspace}
\newcommand{\COMPASSTWO}{\textsc{Compass}-\RM{2}\xspace}
\newcommand{\CERN}{\textsc{Cern}\xspace}
\newcommand{\expi}{exclusive $\pi^0$\xspace}
\newcommand{\phipi}{$\phi_{\pi^{0}}$\xspace}
\newcommand{\lepto}{\textsc{Lepto}\xspace}

\newcommand{\di}{{\rm d}}

\DeclarePairedDelimiter\abs{\lvert}{\rvert}%
\DeclarePairedDelimiter\mean{\langle}{\rangle}%

\begin{document}

\title{Measurement of the exclusive $\pi^0$ muoproduction cross section at COMPASS}
\date{\today}

\author{Matthias Gorzellik}
\email{matthias.gorzellik@cern.ch}
\affiliation{Universit\"at Freiburg, Physikalisches Institut, 79104 Freiburg, Germany}
\collaboration{on behalf of the \COMPASS collaboration}
\noaffiliation

\begin{abstract}

At \COMPASS DVCS and DVMP processes are studied in order to probe the partonic
structure of the nucleon by constraining GPD models.
Extending beyond semi-inclusive deep inelastic scattering, the measurement of
lepton-induced exclusive reactions enables the study of GPDs, which ultimately
reveal the three dimensional picture of the nucleon and the decomposition of
its total angular momentum.
Exploiting the flavour filtering character of DVMP measurements, the \COMPASS
experiment is able to access different combinations of quark and gluon GPDs by
determining the cross sections for various mesons.
We report on the first extraction of the exclusive $\pi^0$ muoproduction cross section in the intermediate $x_{Bj}$ domain ranging from 0.01 to 0.15.
\end{abstract}

\maketitle

\section{Introduction}
Measurements of exclusive leptoproduction of pseudoscalar mesons provide
supplementary data for parameterizations of Generalized Parton Distributions (GPDs).
In the past decade, GPDs have shown to be a very rich construct for both experiments and theory enabling measurements and predictions regarding the inner structure of nucleons.
The GPDs correlate transverse spacial positions and longitudinal momentum fractions and they are related to form factors and parton distribution functions.
There are four parton helicity-conserving (chiral-even) GPDs, denoted by $H^q$, $\tilde{H}^q$, $E^q$, $\tilde{E}^q$ for each quark flavor $q$.
In addition, there are four corresponding GPDs describing  helicity-flip (chiral-odd) processes, $H_T^q$, $\tilde{H}_T^q$, $E_T^q$, $\tilde{E}_T^q$.
While Deeply Virtual Compton Scattering and production of vector mesons are sensitive primarily to $H^q$ and $E^q$, 
at the leading-twist the production of pseudoscalar mesons by longitudinal virtual photons is sensitive to $\tilde{H}^q$ and $\tilde{E}^q$ (related to parton helicity distributions).

Although expected to be suppressed by the inverse of the photon virtuality in the amplitude~\cite{Collins-1996-ID17}, the experimental data on exclusive $\pi^+$ production from HERMES~\cite{hermes1} and on $\pi^0$ production from CLAS~\cite{clas1, clas2} and JLAB Hall A~\cite{hallA1, hallA2} have indicated substantial contributions from transversely polarized virtual photons to the production of spin-0 mesons. 

In the GPD formalism such contributions are possible when a quark helicity-flip GPD couples to a twist-3 meson distribution amplitude~\cite{GK2011, Ahmad-2008-ID20}. In the framework of Ref.~\cite{GK2011}
the pseudoscalar meson production can be described by the following GPDs: $\tilde{H}^q$, $\tilde{E}^q$, $H_T^q$ and $\bar E_{T}^q = 2\tilde{H}_T^q + E_T^q$.
The production of $\pi^+$ or $\pi^0$ mesons exhibit different sensitivities to various GPDs.
When taking into account relative signs and sizes of the GPDs for $u$ and $d$
quarks, and the quark flavour content of the meson, the following differences are expected ~\cite{GK2011}. 
For the $\pi^+$ channel the cross section at small $|t|$, with $t$ being the square of the four momentum transfer to the nucleon, is dominated by the contributions 
from longitudinal virtual photons, of which a major part comes from the pion pole exchange (the main contributor to $\tilde{E}$). Besides, the contributions 
from $\tilde{H}$ and $H_T$ are significant, and a strong cancellation between 
$\bar E_{T}$ for $u$ and $d$ quarks occurs.
On the contrary, the pion pole exchange is absent in the $\pi^0$ case, contributions from $\tilde{H}$ and $H_T$ are small and a large contribution from transversely polarized photons is generated mainly by $\bar E_{T}$.

These differences are manifested by a different magnitude of the cross section and different kinematic dependencies for each channel, in particular, for the predicted $t$-dependencies at small $\abs{t}$. 
While for $\pi^+$ a fast decrease of the cross section with increasing $|t|$ is predicted by theoretical models and confirmed by the experimental results from HERMES~\cite{hermes1}, a dip is expected as $|t| \rightarrow 0$ for the $\pi^0$ channel~\cite{GK2011} and confirmed in the large-$x$ domain by the recent JLAB results~\cite{clas1,clas2,hallA2}.

It should be noted that the only constrains for modelling $\bar E_{T}$ come from a lattice-QCD study~\cite{QCDSF} of its moments. Therefore \COMPASS
measurements of exclusive $\pi^0$ production may provide a new input for modelling transversity GPDs in general, and in particular for an 'elusive' 
$\bar{E}_{T}$. \\

The unpolarized reduced meson production cross section reads
\begin{align*}
\frac{\di^2 \sigma ^{\gamma^* p }}{ \di t \di \phi_{\pi^{0}}} & = \frac{1}{2\pi}
\Big[\frac{\di \sigma_{T}}{\di t} + \epsilon \frac{\di \sigma_{L}}{\di t}
+ \epsilon \cos \left( 2\phi_{\pi^{0}} \right) \frac{\di \sigma_{TT}}{\di t}\\
\nonumber
&+ \sqrt{2 \epsilon \left( 1 + \epsilon \right)} \cos \left( \phi_{\pi^{0}} \right) 
\frac{\di \sigma_{LT}}{\di t}\Big], &
\end{align*}
where $\sigma_{T}$, $\sigma_{L}$, $\sigma_{TT}$, $\sigma_{LT}$ are the structure functions, $\epsilon$ is the virtual photon polarization parameter and \phipi represents the angle between the leptonic and hadronic planes (Trento convention~\cite{Bacchetta-2004-ID25}). Here, the subscript T(L) denotes the contribution from transversely (longitudinally) polarized virtual photons, while the subscripts TT and LT denote the contributions from the interference between transversely-transversely and longitudinally-transversely polarized virtual photons.
According to Ref. \cite{GK2011, clas1}, the structure functions are connected to convolutions of GPDs with the elementary process (denoted by brackets), via the following relations:
\begin{align*}
    \frac{\di \sigma_{T}}{\di t} &\propto \Big[ (1-\xi^2) \abs{\mean{H_T}}^2 - \frac{t'}{8m^2}\abs{\mean{\bar E_{T}}}^2 \Big],\\
    \frac{\di \sigma_{L}}{\di t} &\propto 
    \Big[
        (1-\xi^2) \abs{\mean{\tilde{H}}}^2\\
        & - 2\xi^2 Re\left[ \mean{\tilde{H}}^\ast\mean{\tilde{E}} \right]
        - \frac{t'}{4m^2}\xi^2\abs{\mean{\tilde{E}}}^2
    \Big],\\
    \frac{\di \sigma_{TT}}{\di t} &\propto t' \abs{\mean{\bar E_T}}^2,\\
    \frac{\di \sigma_{LT}}{\di t} &\propto \xi \sqrt{1-\xi^2}\sqrt{-t'} Re\left[ \mean{H_T}^\ast\mean{\tilde{E}} \right],
\end{align*}
with $t' = t - t_{min}$, $t_{min}$ being the smallest possible square of four momentum transfer and $m$ being the mass of the proton.

\section{Experimental setup}
\COMPASS is a fixed target experiment located at \CERN with a tertiary $\mu^+$ or $\mu^-$ beam that is focused on a 2.5\,m long unpolarized liquid hydrogen target. The polarization of $\pm$ 80\,\% changes with the beam charge and the average beam momentum is 160\,GeV/c with a spread of 6\,GeV/c.
\COMPASS uses an open field, two stage spectrometer with a large variety of different tracking detectors for the reconstruction of charged tracks. A muon system and a ring imaging Cherenkov counter allow for particle identification. Energies are measured with hadronic or electromagnetic calorimeters, respectively.
Within the \COMPASSTWO program, the spectrometer was complemented with an additional electromagnetic calorimeter positioned straight after the target, together with a target time-of-flight system surrounding the target that allows the detection of recoiling protons.\\
The presented analysis is based on a data set that was recorded during a four weeks pilot run conducted 2012.

\section{Event selection}
The event selection exploits the overconstrained kinematic of the \expi production process 
\begin{align*}
\mu p\rightarrow \mu'p'\pi^0\rightarrow \mu'p'\gamma\gamma.
\end{align*}
Events with one identified outgoing muon, at least two clusters in the electromagnetic calorimeter and at least one recoiling proton track, are examined. The combinatorial ambiguity is resolved by applying cuts on variables that are sensitive on the exclusivity of the event (see Fig. \ref{fig:dvmp} for the definition of the four momenta):
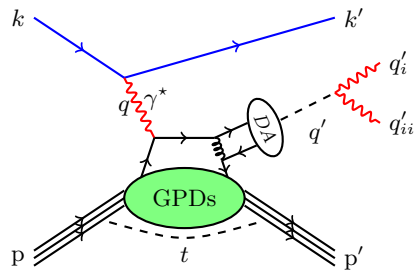
\begin{figure}[ht]
    \centering
    \begin{tikzpicture}[
            thick,
            scale=0.4,
            level/.style={level distance=1.5cm},
            level 2/.style={sibling distance=2.6cm},
            level 3/.style={sibling distance=2cm}
        ]
    
    	\path[draw, lepton] (0,8) -- (3,6) node[pos=0, left] {$k$};
    
    	\path[draw, lepton] (3,6) -- (10,8) node[pos=1, right] {$k'$};
    
    	\path[draw, photon] (3,6) -- (4,4) node[pos=0.35, right] {$\gamma^{\star}$} node[pos=0.5, left] {$q$};
    
    	\path[draw, nucleon] (4,4) -- (6,4);
    	\path[draw, nucleon] (3.5,2.5) -- (4,4);
    	\path[draw, nucleon] (6.25,3.25) -- (6.5,2.5);
    	\path[draw, gluon] (6,4) -- (6.25,3.25);
    	
    	\path[draw, nucleon] (6,4) -- (7.1,4.5);
    	\path[draw, nucleon] (7.4,3.75) -- (6.25,3.25);
    	\draw[rotate around={20:(7.75,4.9)}] (7.5,4.4) ellipse (.5 and 1) node[rotate=-65] {\scriptsize{\textit{DA}}};
    	\path[draw, dashed] (8.2,4.6) -- (10,5.5) node[pos=0.35, below right] {$q'$};
    	\path[draw, photon] (10,5.5) -- (11.5,6.5) node[pos=1, right] {$q'_i$};
    	\path[draw, photon] (10,5.5) -- (11.5,4.5) node[pos=1, right] {$q'_{ii}$};

    	\path[draw, nucleon] (0,0.25) -- (3,2.25);
    	\path[draw, nucleon] (0,0) -- (3,2) node[pos=0, left] {$\mathrm{p}$};
    	\path[draw, nucleon] (0,-0.25) -- (3,1.75);
    
    	\path[draw, nucleon] (7,2.25) -- (10,0.25);
    	\path[draw, nucleon] (7,2) -- (10,0) node[pos=1, right] {$\mathrm{p}'$};
    	\path[draw, nucleon] (7,1.75) -- (10,-0.25);
    
    	\draw[fill=green!50] (5,2) ellipse (2 and 1) node {GPDs};
    
    	\draw[draw=black, dashed] (2.5,1.2) .. controls (5,0.5) .. (7.5,1.2) node[pos=0.5, below] {$t$};
    
    \end{tikzpicture}
    \caption{Schematic diagram for the \expi production process.}
    \label{fig:dvmp}
\end{figure}
\begin{itemize}
    \item four momentum balance,\\ $M^2_{X} = (k+p-k'-q'-p')^2$,
    \item transverse momentum balance,\\ $\Delta p_{\bot} = p'_{\bot,spec.} - p'_{\bot,meas.}$,
    \item coplanarity,  $\Delta \varphi = \varphi_{spec.} - \varphi_{meas.}$, where $\varphi$ denotes the azimuthal angle of the proton in the lab,
    \item vertex pointing, $\Delta z = z_{interp.} - z_{A}$.
    \item mass of two $\gamma$ system, $M_{\gamma\gamma} = (q'_{i} + q'_{ii})^2$,
\end{itemize}
Here, the subscript \textit{meas} denotes a quantity measured by the recoil proton detector and \textit{spec} denotes a quantity derived from spectormeter measurements only, assuming exclusivity $p'_{spec.} = k+p-k'-q'$.
The vertex pointing variable checks the compatibility of the longitudinal position of the vertex and the proton track. 
This is achieved by comparing the measured hit position in the inner ring of the detector $z_{A}$
to a predicted hit position $z_{interp.}$, found by interpolating between the interaction vertex and the hit in the outer ring of the detector.
\begin{figure}[ht]
    \centering
    \includegraphics[width=.49\linewidth]{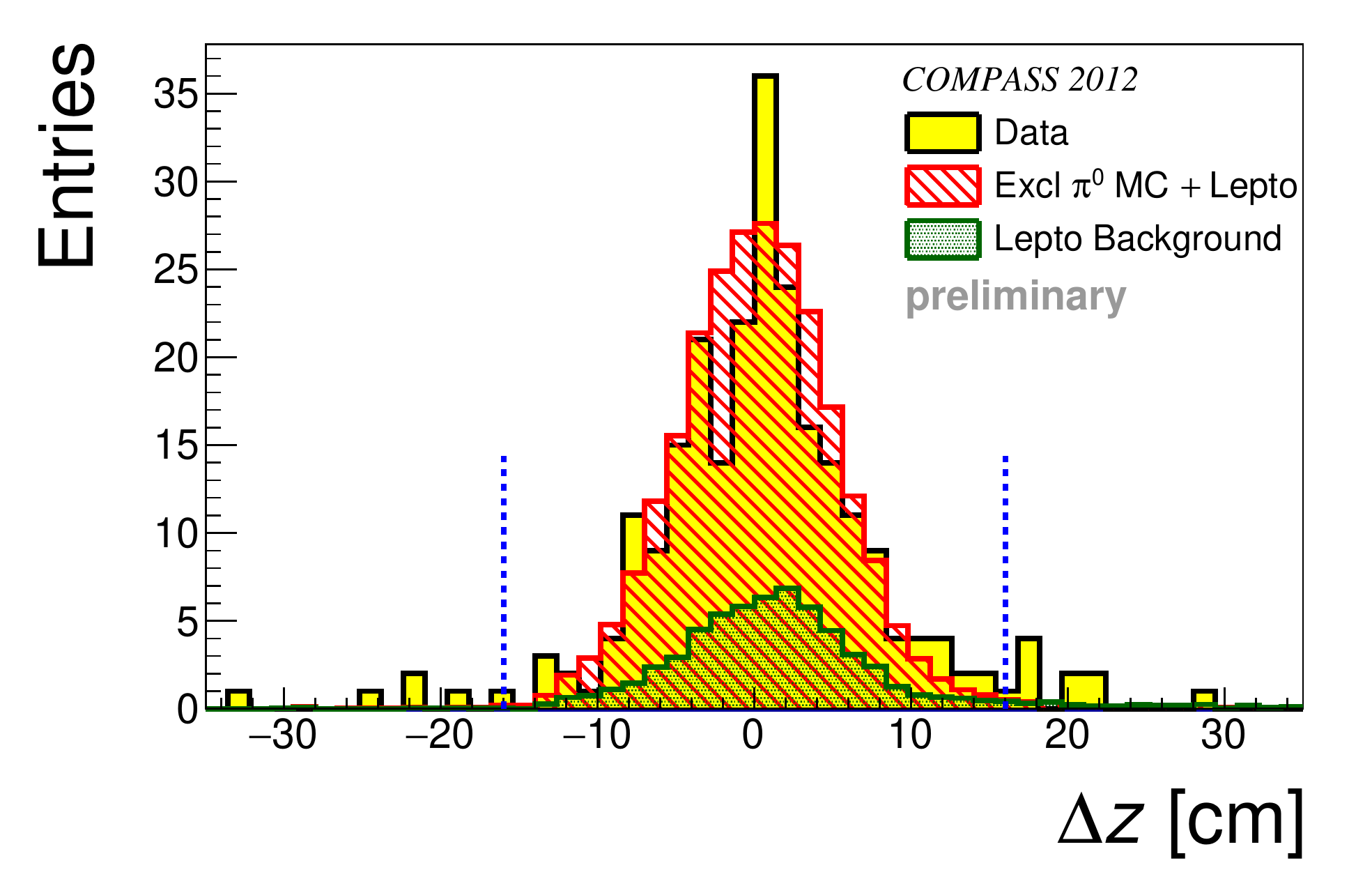}
    \includegraphics[width=.49\linewidth]{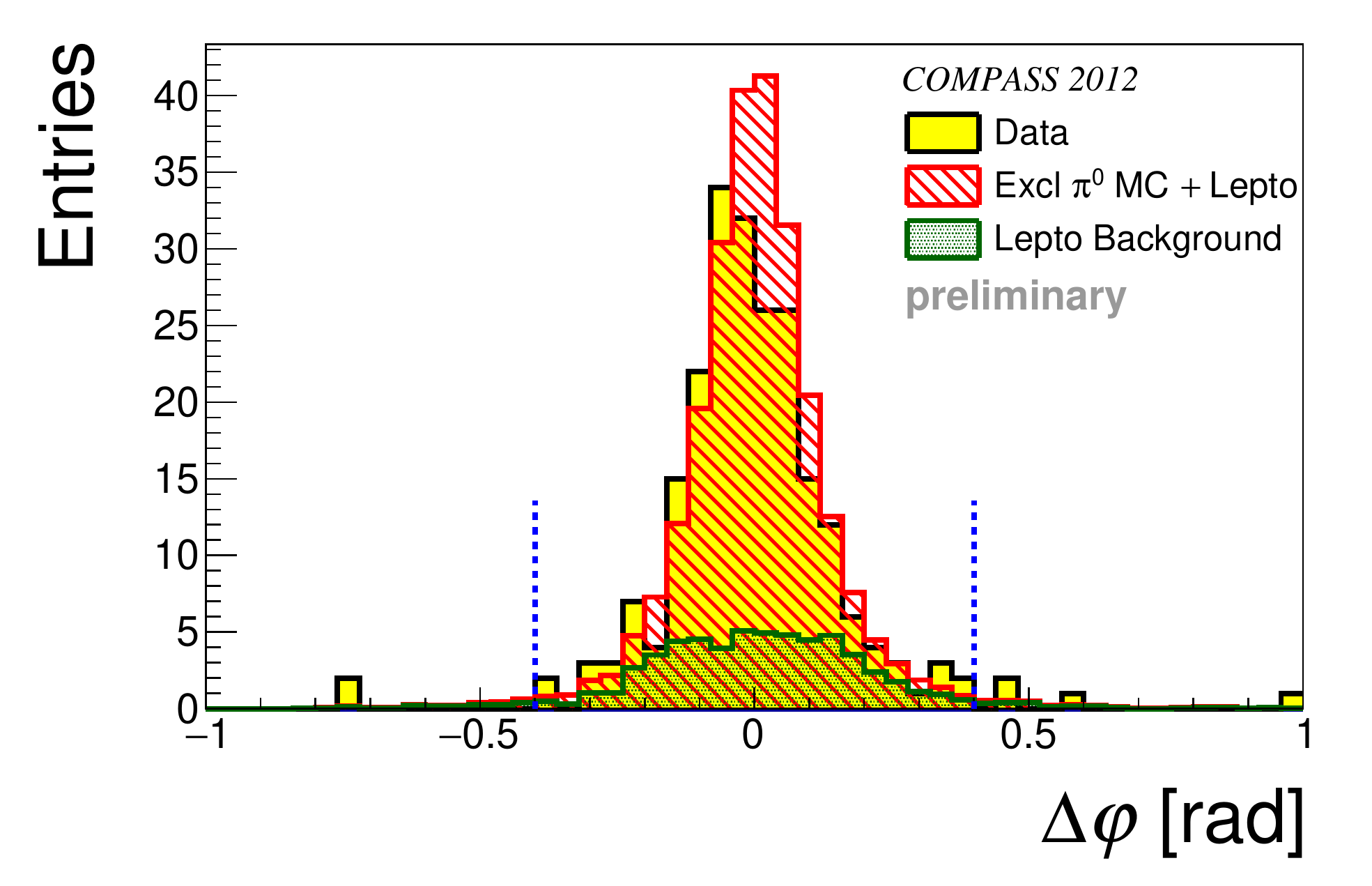}\\
    \includegraphics[width=.49\linewidth]{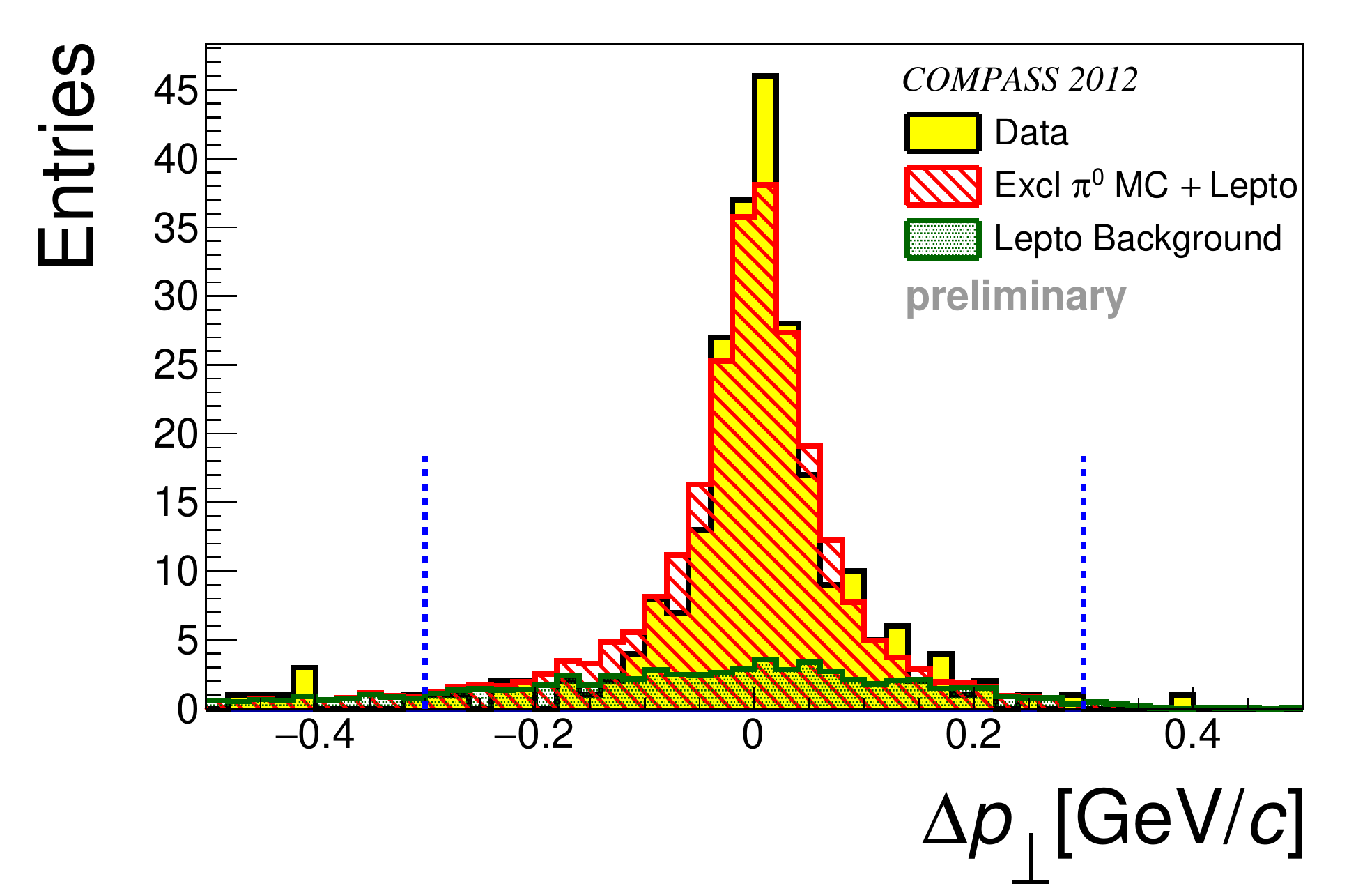}
    \includegraphics[width=.49\linewidth]{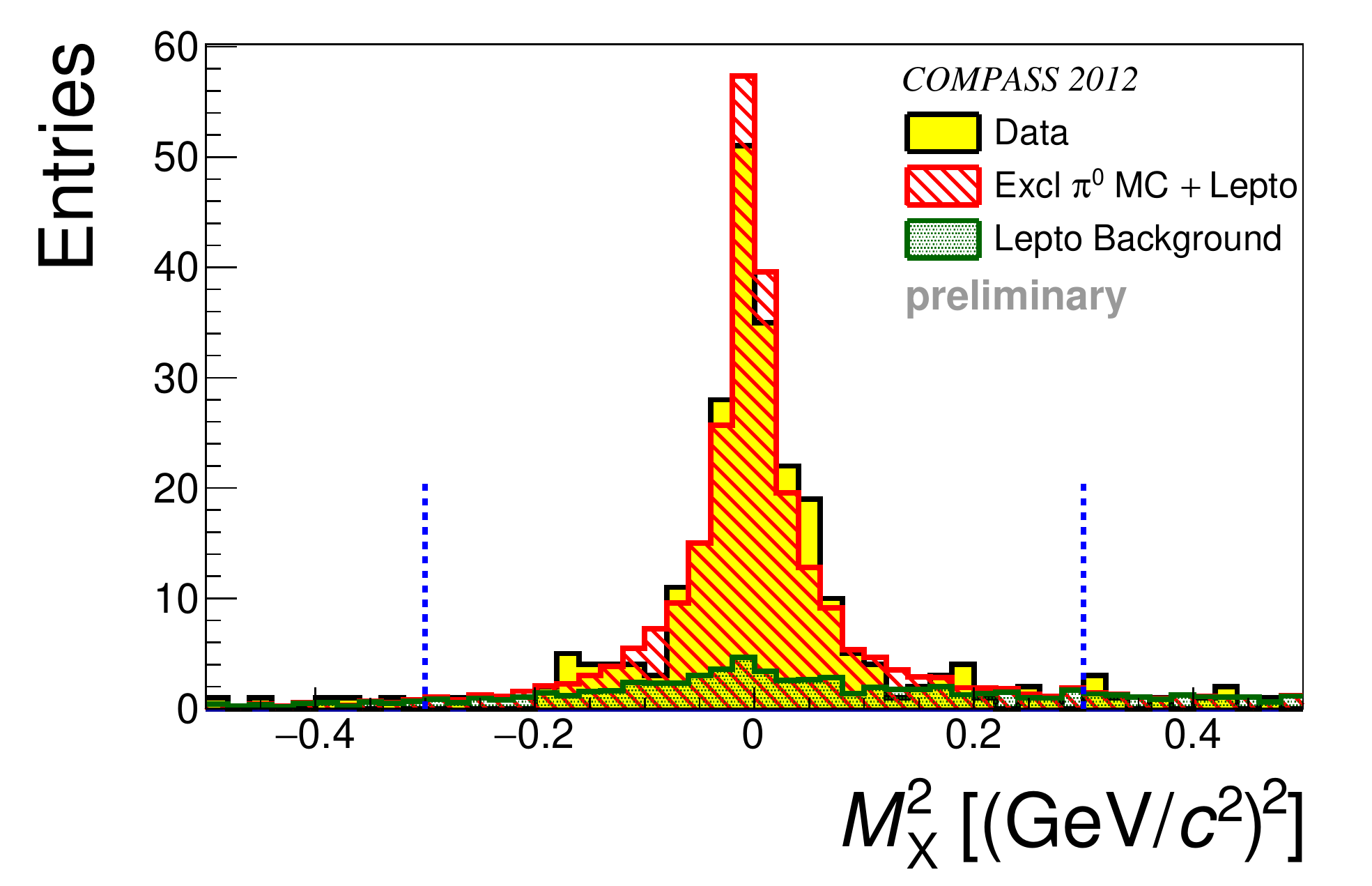}
    \caption{Variables sensitive on the exclusivity of the event, used in the cuts of the event selection. The blue dotted lines indicate the applied cuts.}
    \label{fig:excl_vars}
\end{figure}
Finally, combinatorial unambiguity is required and a cut on the mass of the two $\gamma$ system is performed.
\begin{figure}[ht]
    \centering
    \includegraphics[width=\linewidth]{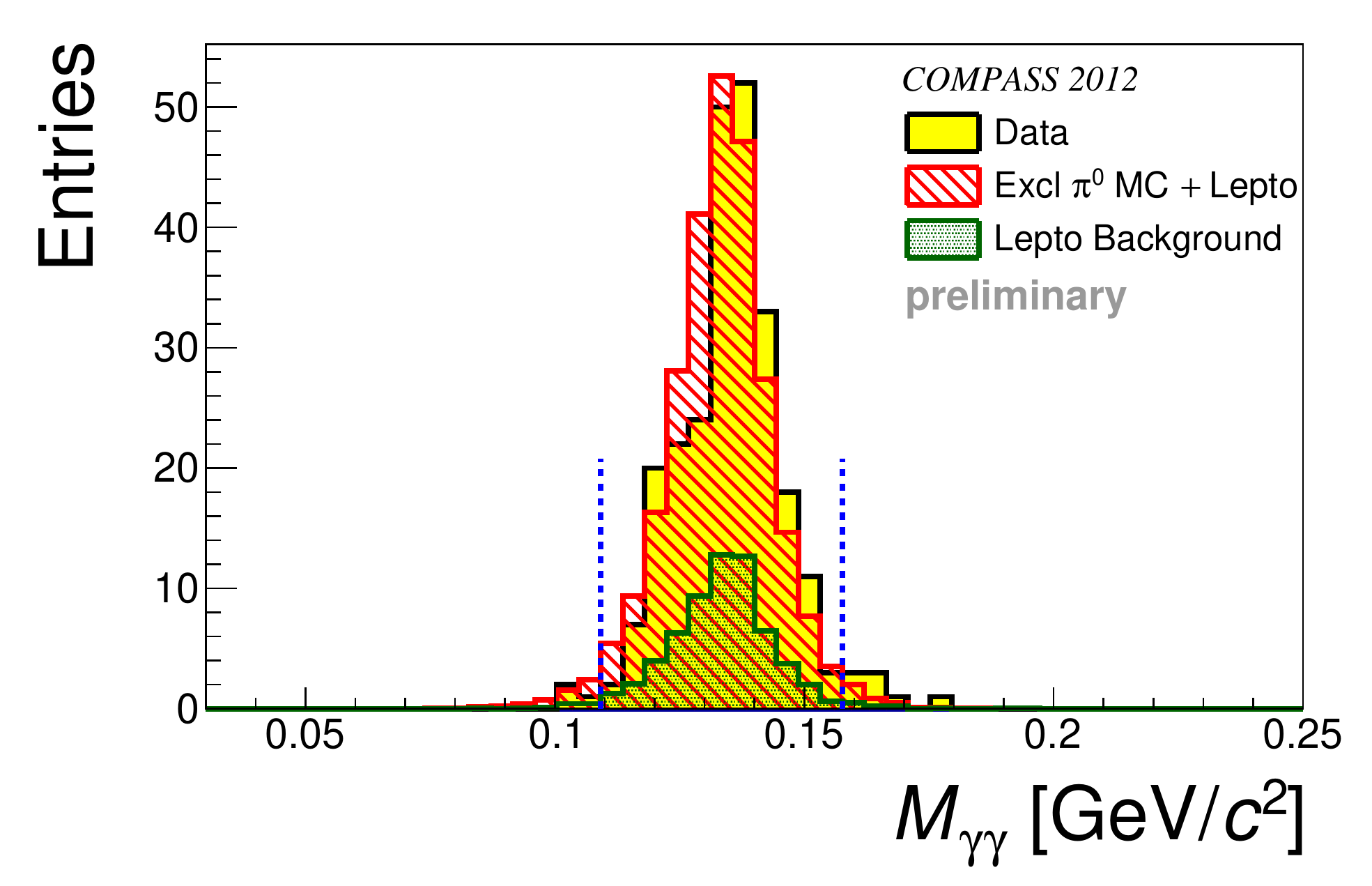}
    \caption{Invariant mass of the two $\gamma$ system. The blue dotted lines indicate the applied cut.}
    \label{fig:mass}
\end{figure}

The background is further reduced by the application of a kinematically constrained fit. The fit is fed with all measured quantities, their uncertainties and correlations, and performs a chi-squared minimization of the event with constraints on energy and momentum conservation. In addition, the mass of the two $\gamma$ system is constrained to the $\pi^0$ PDG mass. Cuts are applied on pull distributions of selected variables, which allow for a good control over each individual cut.
\begin{figure}[ht]
    \centering
    \includegraphics[width=\linewidth]{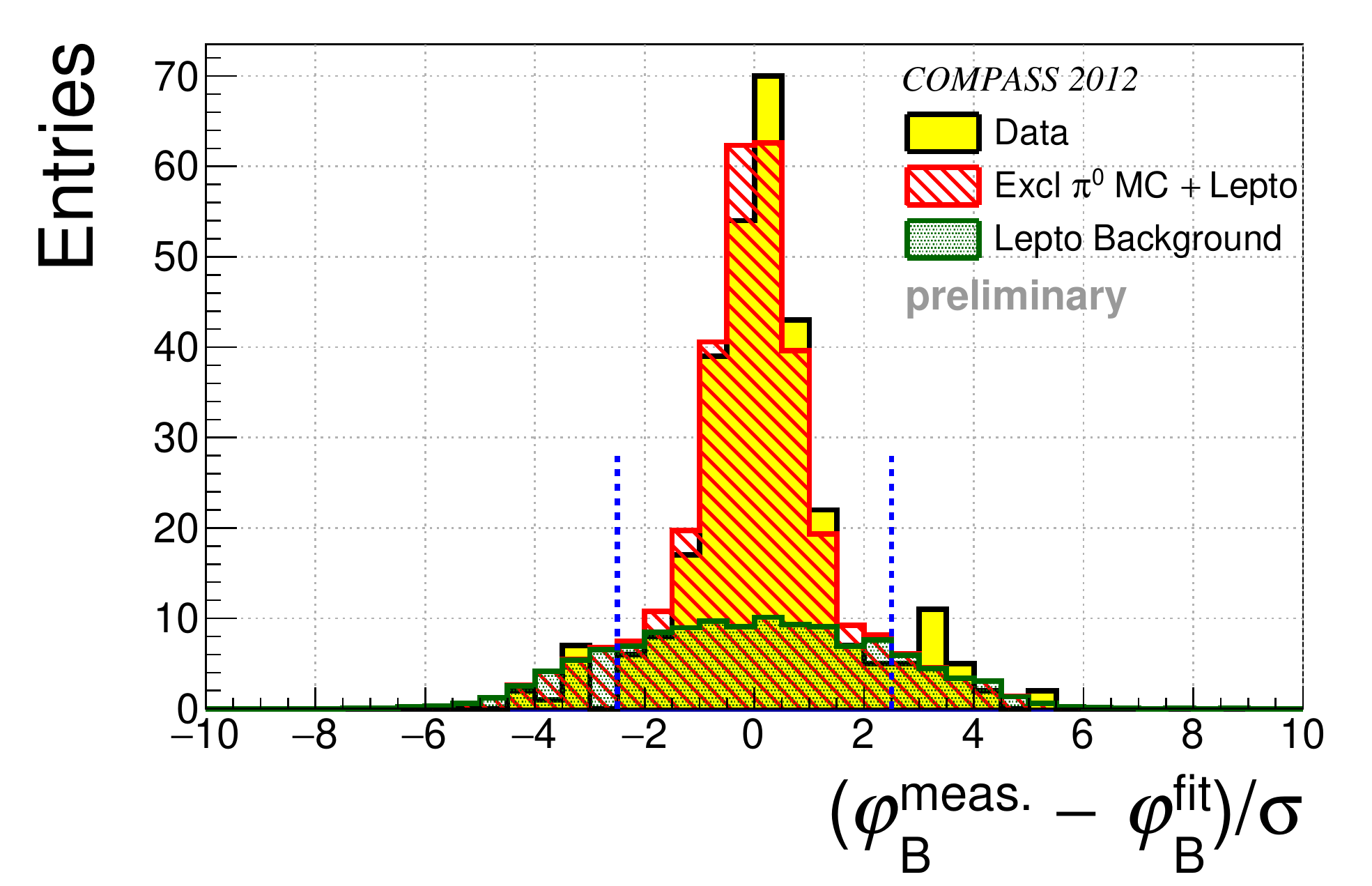}
    \caption{Pull distribution for the azimuthal angle of the proton in the outer ring of the detector. The blue dotted lines indicate the applied cut.}
    \label{fig:pull}
\end{figure}
In the following, the kinematic variables determined by the kinematic fit are used since the fit offers their most precise determination.

\section{Results}
A complete and detailed simulation of the experimental setup is used to build the acceptance of the experiment. To determine the background originating from non-exclusive events, the data is modeled with Monte Carlo in both kinematic regions where the signal dominates and regions where the background dominates.
Here, \lepto 6.5.1 is used for the non-exclusive portion. For the signal contribution a parametrization from Goloskokov and Kroll is used together with a dedicated Monte Carlo \cite{1207.0333}. This background is then subtracted from the data. Other sources of background, like miss-identified exclusive $\omega \rightarrow \gamma\gamma\gamma$ where one $\gamma$ is lost, are found to be negligible. The range of the kinematic variables and their mean values are listed in Tab. \ref{tab:kin}.\\
\begin{table}[ht]
    \centering
    \caption{Kinematic range covered in the analysis and mean values.}
    \label{tab:kin}    
    \begin{tabular}{c|r|r|r}
         & lower limit & upper limit & mean\\
         \toprule
        $Q^{2}/(\text{GeV}/c)^{2}$ & 1.0 & 5.0 & 2.0\\
        $\nu/\text{GeV}$ & 8.5 & 28.0 & 12.8 \\
        $\abs{t}/(\text{GeV}/c)^{2}$ & 0.08 & 0.64 & 0.256
    \end{tabular}
\end{table}

\begin{figure}[ht]
    \centering
    \includegraphics[width=\linewidth]{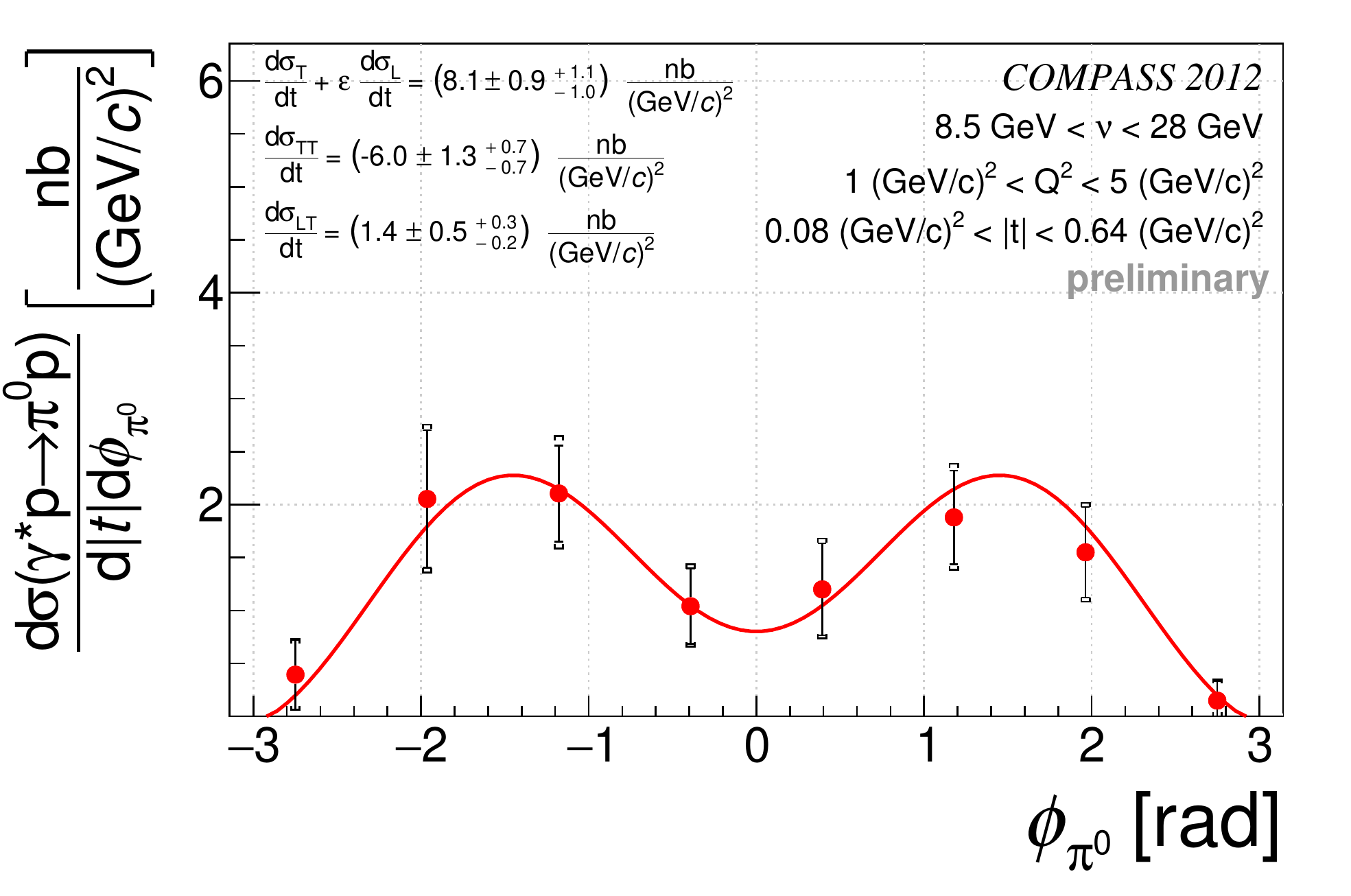}
    \caption{The \expi cross section as a function of \phipi.}
    \label{fig:x_sec_phi}
\end{figure}

We extract the cross section as a function of \phipi in one bin of $t$ and eight equidistant bins in \phipi.
For the definition of the photon flux we use the Hand convention~\cite{Hand-1963-ID24}.
The red dots in Fig. \ref{fig:x_sec_phi} show the measured cross section for each bin, a binned maximum likelihood fit is used to extract the amplitudes of the modulations (red curve). The values of the fitted parameters are found to be
\newcommand{\unit}{\,\frac{\text{nb}}{(\text{GeV}/c)^{2}}}
\begin{align*}
    \frac{d\sigma_{T}}{dt} + \varepsilon \frac{d\sigma_{L}}{dt} &= 8.1\pm0.9 ^{+1.1}_{-1.0} \unit,\\
    \frac{d\sigma_{TT}}{dt} &= -6.0\pm1.3 ^{+0.7}_{-0.7} \unit,\\
    \frac{d\sigma_{LT}}{dt} &= 1.4\pm0.5 ^{+0.3}_{-1.2} \unit.
\end{align*}

\begin{figure}[ht]
    \centering
    \includegraphics[width=\linewidth]{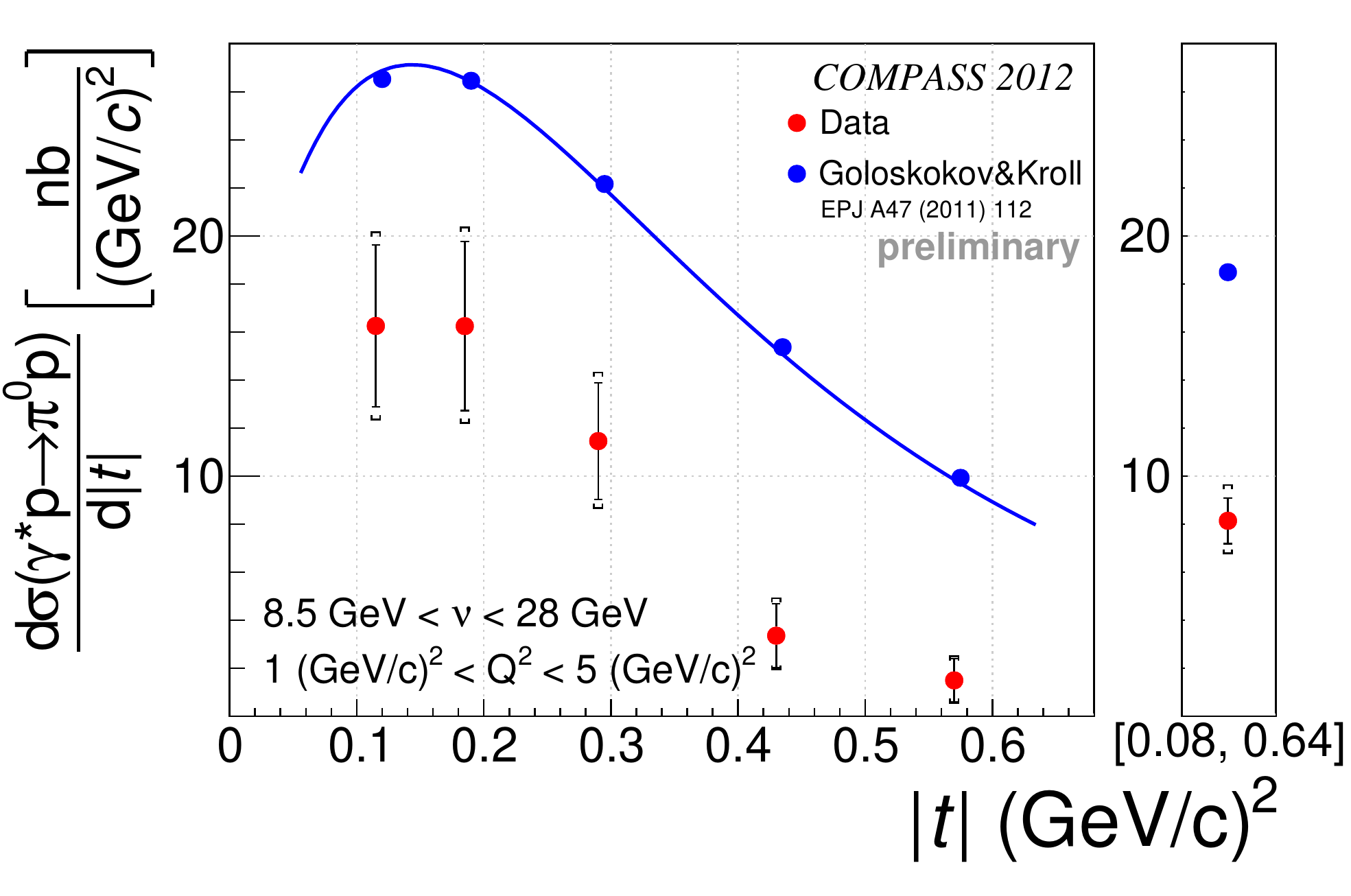}
    \caption{Left panel: \expi cross section as a function of $t$. Right panel: mean cross section for the whole range in $t$.}
    \label{fig:x_sec_t}
\end{figure}

After integration in \phipi, we extract the cross section in bins of $t$. 
In addition, we extract the mean cross section for the full $t$ range, displayed in the right panel of Fig. \ref{fig:x_sec_t}.
In Fig. \ref{fig:x_sec_phi} and Fig. \ref{fig:x_sec_t}, the inner error bars represent the statistical uncertainty, while the outer ones represent the square root of the quadratic sum of statistical and systematic uncertainties.\\
The systematic uncertainty in this measurement is primarily given by scaling which can be decomposed into effects entering through the absolute normalization and background estimation as well as from threshold effects and kinematic uncertainties.

Being the first measurement in its kinematic domain, it is worth comparing it to model predictions.
Using the model of Goloskokov and Kroll, the blue curve in Fig. \ref{fig:x_sec_t} shows the integrated cross section while the blue dots show the mean integrated cross section for the particular bin obtained from the model.
When comparing the magnitude of the cross section, we observe that the prediction overshoots our measurement by approximately a factor of two, which highlights our measurement to be a valuable input parameter for model parameterizations.\\
Because of the limited statistics, we do not favor any model \cite{GK2011, Ahmad-2008-ID20} for the behaviour of the cross section at small $\abs{t}$.\\
Looking at the cross section as a function of \phipi, we observe a large contribution from $\sigma_{TT}$ and a small, positive contribution from $\sigma_{LT}$. This is again an indication that the cross section is especially driven by the transversely polarized photons.\\

Presently, the \COMPASS collaboration takes more data for this interesting reaction, which will increase the statistics by a factor of approximately 15. This will enable us to study the evolution of the \phipi modulations in bins of $t$ and with the increased statistics, it will also allow us to study the $t$ dependant cross section in the range close to zero.\\

\begin{acknowledgments}
The author acknowledges financial support by the German Bundesministerium für Bildung und Forschung (BMBF) and by the DFG
Research Training Group 2044 ‘Mass and Symmetries after the Discovery of the Higgs Particle at the LHC’.
\end{acknowledgments}

\bibliography{main}{}
\bibliographystyle{apsrev4-1}
\end{document}